\documentclass{revtex4}
\usepackage{graphicx}
\def\ra{\rightarrow}
\def\to{\ra}

\def\be{\begin{equation}}
\def\ee{\end{equation}}
\def\bea{\begin{eqnarray}}
\def\eea{\end{eqnarray}}

\def\wt{\widetilde}
\def\epem{e^+e^-}
\def\bb{b\anti{b}}

\def\bbA{\bb \ha}
\def\tanb{\tan\beta}
\def\cotb{\cot\beta}
\def\sina{\sin\alpha}
\def\cosa{\cos\alpha}
\def\sinb{\sin\beta}
\def\cosb{\cos\beta}

\def\eps{\epsilon}
\def\hl{h}
\def\hh{H}
\def\ha{A}
\def\hp{H^+}
\def\hm{H^-}
\def\hpm{H^\pm}
\def\mhpm{m_{\hpm}}
\def\anti{\overline}
\def\mhh{m_{\hh}}
\def\mha{m_{\ha}}
\def\mhl{m_{\hl}}
\def\gev{~{\rm GeV}}
\def\tev{~{\rm TeV}}
\def\fbi{~{\rm fb}^{-1}}
\def\fb{~{\rm fb}}
\def\call{{\cal L}}
\def\rts{\sqrt s}

\def\vev#1{\langle #1 \rangle}

\def\gamhatot{\Gamma_{\rm tot}^{\ha}}
\def\gamhhtot{\Gamma_{\rm tot}^{\hh}}

\def\lsim{\mathrel{\raise.3ex\hbox{$<$\kern-.75em\lower1ex\hbox{$\sim$}}}}
\def\gsim{\mathrel{\raise.3ex\hbox{$>$\kern-.75em\lower1ex\hbox{$\sim$}}}}
\def\ifmath#1{\relax\ifmmode #1\else $#1$\fi}

\begin{document}
\bibliographystyle{revtex}
\begin{titlepage}
\def\thefootnote{\fnsymbol{footnote}}       

\begin{center}
\mbox{ } 

\end{center}
\vskip -1.0cm
\begin{flushright}
\Large
\vspace*{-2cm}
\mbox{\hspace{10.2cm} hep-ph/0112334} \\
\mbox{\hspace{12.0cm} December, 2001}
\end{flushright}
\begin{center}
\vskip 3.0cm
{\Huge\bf
Determination of \boldmath$\tanb$
\smallskip
at a Future \boldmath$e^+e^-$ Linear Collider
}$^\dagger$
\vskip 1cm
{{\LARGE\bf J. Gunion$^1$, T. Han$^2$, J. Jiang$^3$, S. Mrenna$^4$, A. Sopczak$^5$}\\
\smallskip
\Large 
$^1$Davis Institute for HEP, U. of California, Davis, CA \\
$^2$Dept. of Physics, U. of Wisconsin, Madison, WI\\
$^3$Argonne National Laboratory, Chicago, IL \\
$^4$Fermilab, Batavia, IL \\
$^5$Lancaster University, UK}

\vskip 2.5cm
\centerline{\Large \bf Abstract}
\end{center}

\begin{center}
\begin{minipage}{16cm}
\large
\renewcommand{\baselineskip}{10pt}
It is widely stated that
the ratio of neutral Higgs field vacuum expectation values, $\tanb$,
is one of the most difficult parameters to determine
in either the Minimal Supersymmetric Standard Model (MSSM)
or a general type-II Two-Higgs Doublet Model (2HDM).
Assuming an energy and integrated
luminosity of $\rts=500\gev$ and $\call=2000\fbi$,
we show that a very accurate determination of $\tanb$ 
will often be possible using Higgs production rates and/or Higgs decays.
Based on a TESLA simulation, and
assuming no other light Higgs bosons and $100\leq \mha\leq 200\gev$, 
we find that the rate for the process $\epem\to\bb\ha\to\bb\bb$ 
provides an excellent determination
of $\tanb$ at high $\tanb$. In the MSSM
Higgs sector, the rate for $\epem\to\bb\ha+\bb\hh\to \bb\bb$
($\epem\to \hh\ha\to\bb\bb$) provides a good determination
of $\tanb$ at high (low) $\tanb$, respectively, at moderate $\mha$ values.
We also show that direct measurement of the average total width
of the $\hh$ and $\ha$ in  $\epem\to\hh\ha\to \bb\bb$ events provides
an excellent determination of $\tanb$ at large $\tanb$.

\vspace{1cm}

{\sl 
Contribution to the 
Snowmass 2001 Workshop on ``The Future of Particle Physics'', Snowmass, CO, USA, July 2001
\vspace{-3cm}
}

\end{minipage}
\vfill \flushleft{$^\dagger$ \small Work supported in part by the U.S.
  Department of Energy, the Davis Institute for Particle Physics and
  the Wisconsin Alumni Research Foundation.}
\end{center}

\end{titlepage}

\newpage

\vspace*{0.2cm}

\title{Determination of \boldmath$\tanb$
at a Future \boldmath$e^+e^-$ Linear Collider}

\author{J. Gunion$^1$, T. Han$^2$, J. Jiang$^3$, S. Mrenna$^4$, A. Sopczak$^5$}
\affiliation{\vspace*{0.1in}
$^1$Davis Institute for HEP, U. of California, Davis, CA \\
$^2$Dept. of Physics, U. of Wisconsin, Madison, WI\\
$^3$Argonne National Laboratory, Chicago, IL \\
$^4$Fermilab, Batavia, IL \\
$^5$Lancaster University, UK}

\vspace*{-1.7cm}
\begin{abstract}
It is widely stated that
the ratio of neutral Higgs field vacuum expectation values, $\tanb$,
is one of the most difficult parameters to determine
in either the Minimal Supersymmetric Standard Model (MSSM)
or a general type-II Two-Higgs Doublet Model (2HDM).
Assuming an energy and integrated
luminosity of $\rts=500\gev$ and $\call=2000\fbi$,
we show that a very accurate determination of $\tanb$ 
will often be possible using Higgs production rates and/or Higgs decays.
Based on a TESLA simulation, and
assuming no other light Higgs bosons and $100\leq \mha\leq 200\gev$, 
we find that the rate for the process $\epem\to\bb\ha\to\bb\bb$ 
provides an excellent determination
of $\tanb$ at high $\tanb$. In the MSSM
Higgs sector, the rate for $\epem\to\bb\ha+\bb\hh\to \bb\bb$
($\epem\to \hh\ha\to\bb\bb$) provides a good determination
of $\tanb$ at high (low) $\tanb$, respectively, at moderate $\mha$ values.
We also show that direct measurement of the average total width
of the $\hh$ and $\ha$ in  $\epem\to\hh\ha\to \bb\bb$ events provides
an excellent determination of $\tanb$ at large $\tanb$. 

\end{abstract}

\maketitle

\vspace*{-1.3cm}
\section{Introduction}
\vspace*{-4mm}
A future linear collider has great potential for discovering
new particles and measuring their properties. 
Theories beyond the Standard Model (SM)
that resolve the hierarchy and fine-tuning problems
typically involve extensions of its
single-doublet Higgs sector to at least a 
two-doublet Higgs sector (2HDM). The most attractive such model
is the MSSM, which contains a constrained two-Higgs-doublet
sector. In other cases, the effective
theory below some energy scale is equivalent to a 2HDM extension
of the SM with no other new physics. 
While many parameters of theories beyond the SM
can be measured with high precision, it is often stated
that determination of the important parameter 
$\tanb=\vev{H_u^0}/\vev{H_d^0}$ (where $\vev{H_u^0}$
and $\vev{H_d^0}$ are responsible for up-type
quark masses and down-type quark and lepton masses, respectively)
is difficult, especially for large $\tanb$. 
However, Higgs boson couplings are very sensitive to $\tanb$. In particular,
for a CP-conserving Higgs sector we have the following couplings~\cite{Gunion:1989we}
(at tree-level):
\vspace*{-2mm}
\bea
&&\ha\to b\anti b \propto \tanb;\quad \ha\to t\anti t \propto \cotb;\quad
\hp\to t\anti b \propto m_b(1+\gamma_5)\tanb+m_t(1-\gamma_5)\cotb \nonumber\\
&& \hl\to b\anti b \propto -{\sina\over\cosb};\quad \hl\to t\anti t\propto{\cosa\over\sinb};\quad
\hh\to b\anti b\propto {\cosa\over\cosb};\quad \hh\to t\anti t\propto{\sina\over \sinb}\,,
\eea
where $\alpha$ is the mixing angle in the CP-even sector.

In this report, we show how various Higgs boson measurements can
be used to determine $\tanb$, especially when $\tanb$ is large.
Our focus will be on $b\anti b$+Higgs production, Higgs pair production
in the $\bb\bb$ final state and Higgs total widths
as measured in the pair production channel.

\vspace*{-.6cm}
\section{The \boldmath $\bbA\to \bb\bb$ bremsstrahlung process}
\vspace*{-4mm}
The challenge of this study is the low expected production rate and the large 
irreducible background for a four-jet final state, as discussed 
in a previous study~\cite{epj}.
%
Searches for $b\bar b\ha$ and $b\bar b \hl$
were performed in this four-jet channel using 
LEP data taken at the $Z$ resonance~\cite{l3,aleph,delphi,opal}.
%
A LC analysis has been performed using
event generators for 
the signal process $\epem\to\bbA\to\bb\bb$~\cite{generator} and the
$\epem \to eW\nu,~\epem Z,~WW,~ZZ,~q\anti q~(q=u,d,s,c,b),~t\anti t,~\hl\ha$ 
background processes~\cite{background} that
include initial-state radiation and beamstrahlung.

\begin{figure}[t!]
\caption{\label{fig:effpurity} $b$-tagging performance.}
\begin{center}
\vspace*{-1.5cm}
\includegraphics[width=0.5\textwidth]{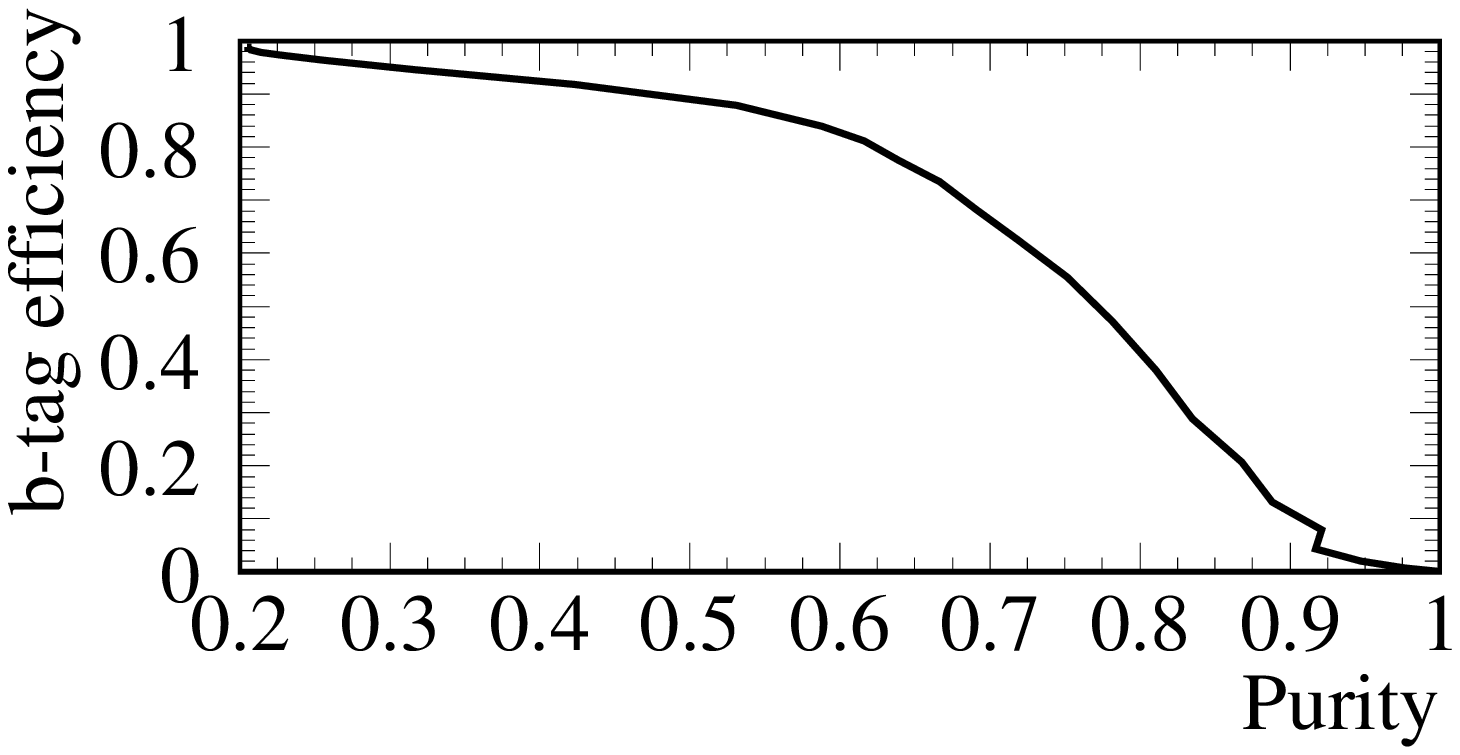}
\vspace*{-0.9cm}
\end{center}
\end{figure}

For a 100 GeV pseudoscalar Higgs boson and $\tanb=50$,
the signal cross section is about 
2~fb~\cite{maria,Djouadi:1991tk,Grzadkowski:1999ye}.
The generated events were passed through the fast detector
simulation SGV~\cite{sgv}.
The detector properties closely follow the TESLA detector 
Conceptual Design Report~\cite{tesla}.
The simulation of the $b$-tagging performance is very important
for this analysis.  
The efficiency versus purity distribution for the simulated
b-tagging performance is shown in Fig.~\ref{fig:effpurity} 
for the hadronic event sample $\epem\to q\bar{q}$ for
5 flavors, where
efficiency is the ratio of 
simulated $\bb$ events after the selection
to all simulated $\bb$ events,
and purity is the ratio
of simulated $\bb$ events after the selection
to all selected $q\anti{q}$ events.
Details of the event selection and background reduction are 
described elsewhere~\cite{epj}.

For $\mha=100\gev$ in the 
context of the MSSM, the 
SM-like Higgs boson is the $\hh$ while the light $\hl$
is decoupled from $WW,ZZ$ [$\cos(\beta-\alpha)\sim 1$
and $\sin(\beta-\alpha)\sim 0$]. The $\bb\hl$ coupling
is essentially equal (in magnitude) to the $\bb\ha$ coupling ($\propto
\tanb$ at the tree level) and $\mhl\sim\mha$, implying that
it would not be possible to separate these two signals. 
 Also important will be $\hl\ha$ production, which is 
$\propto\cos(\beta-\alpha)$ and will have full
strength in this particular situation; $\hh\ha$ production will be
strongly suppressed. We focus first on $b\anti b\ha\to\bb\bb$.

\begin{figure}[h!]
\vspace*{-.5cm}
\caption{\label{fig:ida2}
Plots for $\rts=500\gev$ and $\bb\ha$ events only, before including
running of the $b$-quark mass. 
Left: Final background rate vs. signal 
efficiency for $\mha=100\gev$ and $\call=500\fbi$. Right:  
Corresponding $\tanb$ statistical error for $\call=2000\fbi$
and $\mha=100,150,200\gev$.}
\begin{minipage}{0.48\textwidth}
\begin{center}
\vspace*{-1.7cm}
\includegraphics[width=\textwidth]{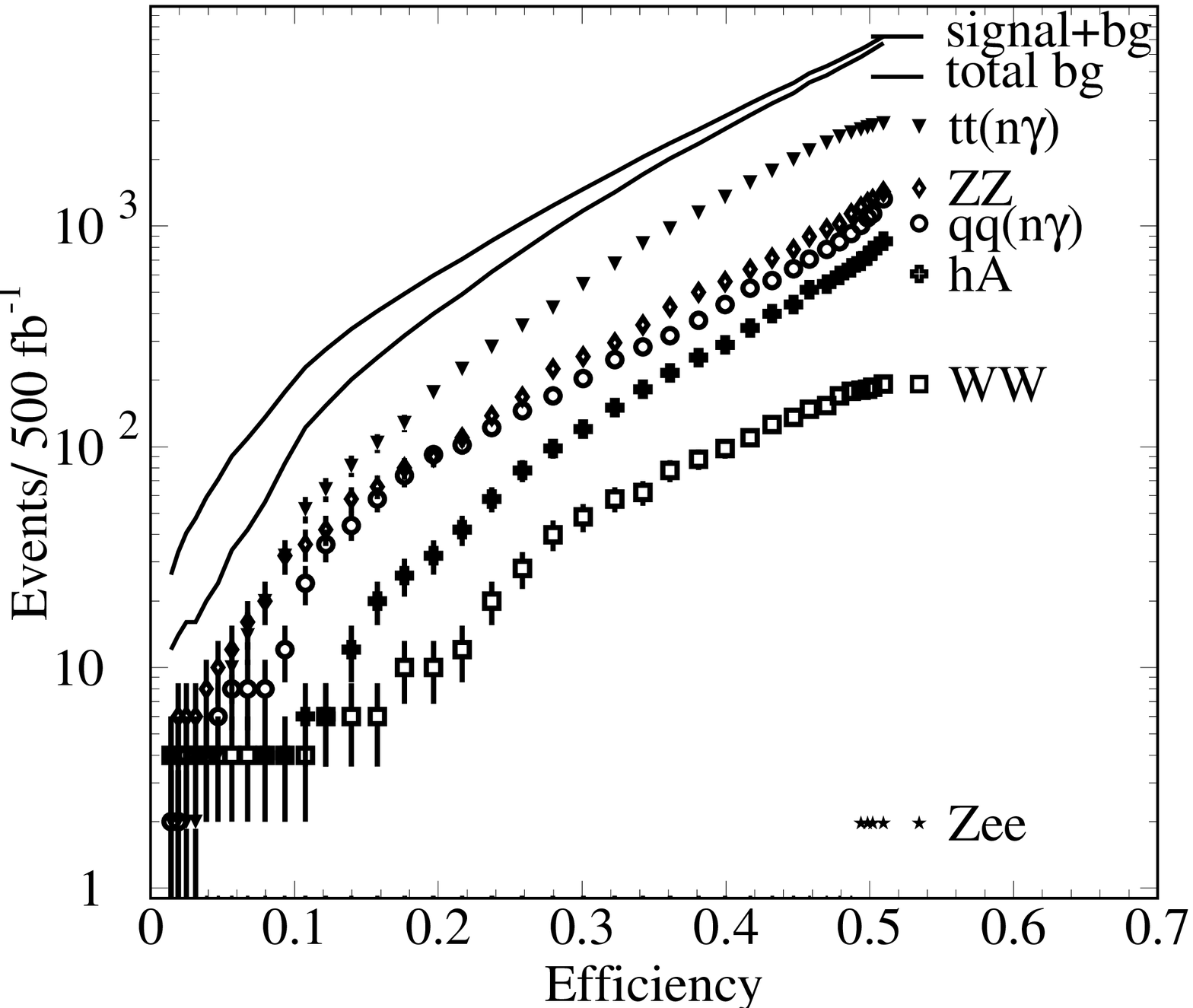}
\end{center}
\end{minipage}
\hfill
\begin{minipage}{0.48\textwidth}
\begin{center}
\includegraphics[width=\textwidth]{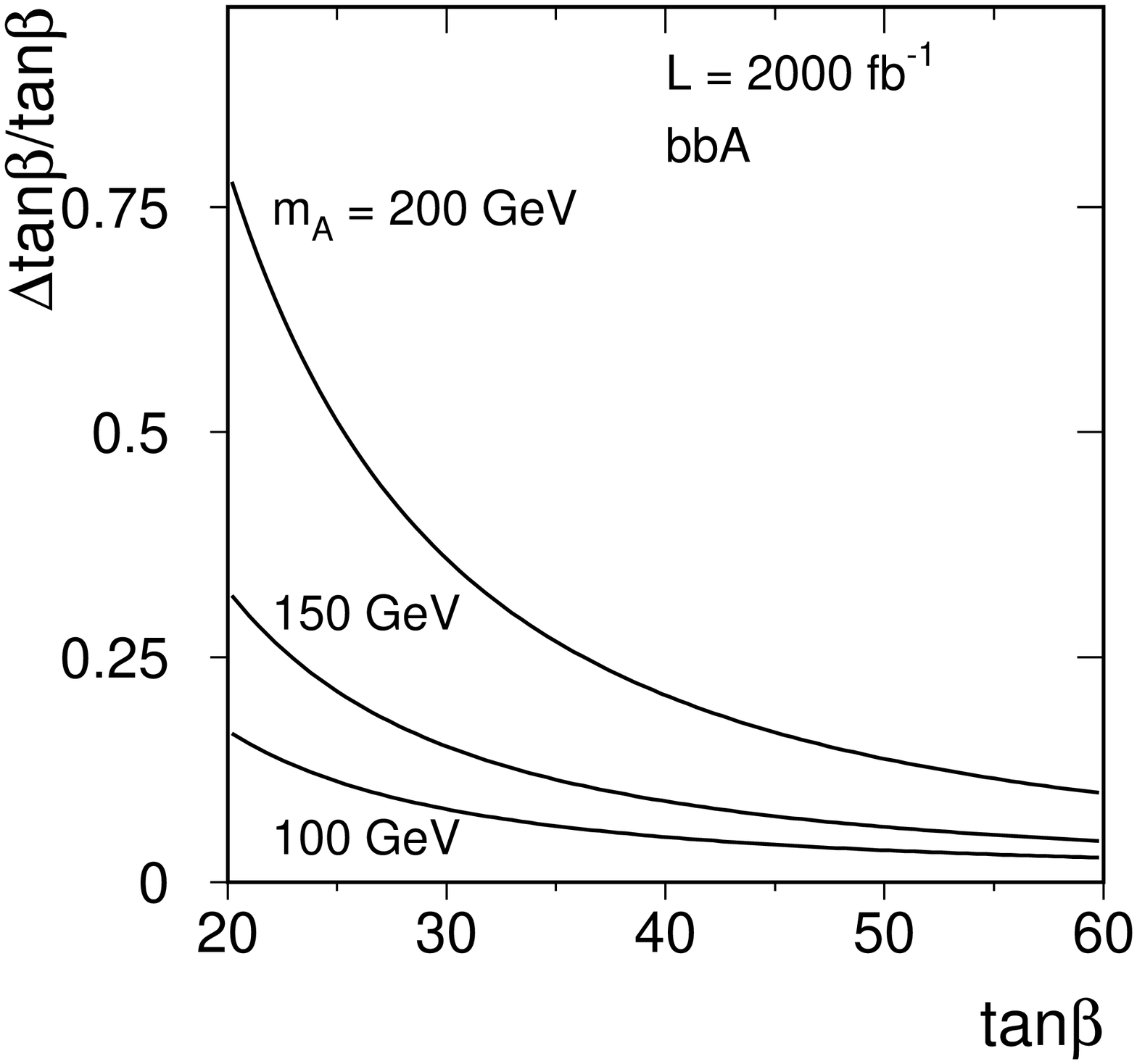}
\end{center}
\end{minipage}
\vspace*{-1cm}
\end{figure}

The expected background rate for a given $b\anti b\ha\to\bb\bb$ signal
efficiency is shown in Fig.~\ref{fig:ida2}. 
One component of the background is $\hl\ha\to \bb\bb$;
our selection procedures are, in part, designed to reduce this piece of the
background as much as possible ({\it e.g.} by removing
events with $m_{\bb}\sim\mhl$ for the second $\bb$ pair).  
Nonetheless, it may lead to 
significant systematic error in the determination of $\tanb$ (see below). 
For the $\bb \ha\to\bb\bb$ signal, 
the sensitivity ${\rm S} / \sqrt{{\rm B}}$
for $\mha=100\gev$ is almost independent 
of the working point choice of signal efficiency in the range 
$\eps_{\rm sel}=5$\% to 50\%.
For a working point choice of 10\% efficiency, 
the total simulated background of about 
16 million events is reduced to 100 background events
with an equal number of signal events at $\tanb=50$.
If this were the only contributing process,
the resulting error on $\tanb= 50$ would be 7\%:
$\Delta\tan^2\beta / \tan^2\beta = \Delta {\rm S} / {\rm S}
=\sqrt{{\rm S} + {\rm B}} / {\rm S} =  0.14.$
For smaller values of $\tanb$, the sensitivity decreases rapidly.
A $5\sigma$ signal detection is still possible for $\tanb= 35$.
In the MSSM context, the $\bb\hl$ signal would essentially double the
number of signal events and have exactly the same $\tanb$ dependence,
yielding $\Delta\tan^2\beta/\tan^2\beta\sim \sqrt{300}/200\sim 0.085$
for $\tanb=50$.

Although 
the number of $\hl\ha$ background events is very small compared to the
other background reactions 
after the event selection,
interference between the signal 
$ \bb\ha\to \bb\bb$ (plus $\bb\hl\to\bb\bb$) and the background 
$ \hl \ha\to \bb\bb$ reaction could be important. 
At the working point, and
after applying the selection procedures, the expected rate for the latter
is $2\pm1$ events for $\call=500\fbi$.
Let us momentarily retain only the $\bb\ha$ signal in discussing 
the interference.
We first calculate the cross sections 
$ \sigma(\epem\to \bb\ha\to \bb\bb)$,
$ \sigma(\epem\to \hl\ha\to \bb\bb)$, and
$ \sigma(\epem\to \bb\ha+\hl\ha\to \bb\bb)$
with CompHEP~\cite{comphep} before selections and define the interference as
$\sigma_{\rm interf}=\sigma_{\bb\ha+\hl\ha}-\sigma_{\bb\ha}-\sigma_{\hl\ha}$.
For the default value $m_{b}=4.62$~GeV, at $\tanb=50$ we obtain 
$\sigma_{\bb\ha}=1.83\pm0.01$~fb, 
$\sigma_{\hl\ha}=36.85\pm0.10$~fb, 
$\sigma_{\bb\ha+\hl\ha}=39.23\pm0.12$~fb, 
$\sigma_{\rm interf}=0.55\pm0.16$~fb.
We observe a constructive interference similar in size to the signal. Thus, 
more signal events are expected than simulated and 
the statistical error estimate is conservative.
After selection cuts, we have found 100 signal events vs. 2 $\hl\ha$
background events.  The maximum interference magnitude arises if 
the interference events are signal-like yielding an interference excess of
$(10+\sqrt 2)^2-100-2\sim 28$, a percentage ($\sim 30\%$) similar to
the ratio obtained before selection cuts. 
If the events from the interference are background-like,
the resulting systematic error will be small, since the $\hl\ha$ background 
is only a small part of the total background. 
Of course, in the MSSM context we have an exact prediction 
as a function of $\tanb$ for
the combined contribution of $\hl\ha\to \bb\bb$ and $\bb\ha\to\bb\bb$
(plus $\bb\hl\to\bb\bb$), including all interferences, 
and this exact prediction can be compared to the data.
In order to test this exact prediction, it may be helpful
to compare theory and experiment for several different event selection
procedures, including ones that give more emphasis to the $\hl\ha$ process.
Of course, this exact prediction depends somewhat on other
MSSM parameters, especially if decays of the $\hl$
or $\hh$ to pairs of supersymmetric particles are allowed or
ratios of certain MSSM parameters are relatively large~\cite{Carena:1998gk}. 
If this type of uncertainty exists,
the systematic error on $\tanb$ can still be controlled by 
simultaneously simulating all sources of $\bb\bb$ events for various 
$\tanb$ values and fitting the complete data set
(assuming that the MSSM parameters are known sufficiently well). 
Another possible theoretical systematic uncertainty derives from 
higher-order corrections. The full NLO QCD corrections 
are given in \cite{Dittmaier:mg,Reina:2001sf}. There it is
found that using the running $b$-quark mass incorporates
the bulk of the NLO corrections. For example,  for $\mha=100\gev$, employing
$m_b(100\gev)\sim 2.92\gev$ vs. $m_b(m_b)\sim 4.62\gev$ yields (before cuts)
a cross section of $\sim 0.75\fb$ vs. $\sim 2\fb$, 
respectively, at $\tanb=50$. 
The signal rates and resulting errors quoted in this section 
are those computed using $m_b=4.62\gev$. Use of the running mass would reduce
the event rates and increase our error estimates;
the resulting errors will be given in the MSSM context in our final figure.
Higher-order corrections of all kinds will be even better known
by the time the Linear Collider (LC) 
is constructed and data is taken and thus should not be
a significant source of systematic uncertainty.
The final source of systematic uncertainty is that associated
with knowing the exact efficiency of the event selection procedure.
At the working point of $\eps_{\rm sel}=10\%$, to achieve 
$\Delta\tanb/\tanb<0.05$ requires $\Delta\eps_{\rm sel}/\eps_{\rm sel}<0.1$,
equivalent to $\Delta\eps_{\rm sel}<1\%$. 
This is probably the best that can be done.

In addition to the $\hl\ha$ Higgs boson background, two other Higgs boson
processes could lead to a $\bb\bb$ topology. 
First, the process $ \epem\to \hh Z$ can give a $\bb\bb$ final state.
In fact, for large $\tanb$ the $\hh Z$ cross section 
is maximal and similar in size to the $\hl\ha$ cross section. Nonetheless,
its contribution to the background is much smaller because
the $ \hh Z\to\bb\bb$ branching is below 10\% compared
to about 80\% for $\hl\ha\to\bb\bb$. 
Since the $\hl\ha$ process contributed only
2\% of the total background, the contribution to the background
from the $\hh Z$ process can be neglected.
The second Higgs boson process leading to a $\bb\bb$ topology is 
that already discussed, $ \epem\to\bb \hl$. The only distinction
between this and the $ \epem\to\bb \ha$ process
is a small difference in the angular distribution
due to the different production matrix elements. 
Thus, the selection efficiency is almost identical. The production rate
of the $ \bb \ha$ process is proportional to 
$\tan^2\beta$ while the $ \bb \hl$
production rate is proportional to $\sin^2\alpha / \cos^2\beta$.
In the MSSM context, this latter factor is $\sim\tan^2\beta$ for 
$\mha=100\gev$ and large $\tanb$ (assuming $M_{\rm SUSY}\sim 1\tev$). 
In the general 2HDM, since $\tanb\approx 1/\cos\beta$ at large $\tanb$, 
the expected rate depends mostly on $\sin\alpha$ and the $\hl$ mass.
In this more general case, if $\mhl \approx \mha$ but 
the MSSM expectation of $\alpha\sim -\beta\sim -\pi/2$ does not hold,
the enhancement of the $\bb\ha$ signal by
the $ \bb \hl$ addition would only allow a determination of
$|\sin\alpha|$ as a function of the presumed value of $\tanb$
(using the constraint that one must obtain the observed
number of $\bb\hl+\bb\ha$ events). Independent measurements
of the $\hh Z$ and $\hl\ha$
 production rates would then be needed to determine
the value of $\beta-\alpha$ and only then could
$\alpha$ and $\beta$ be measured separately.

It is estimated that $\call=2000\fbi$ can be accumulated after several years
of data-taking at the LC. Such high total
luminosity is of particular importance for the $\tanb$ determination.
In Fig.~\ref{fig:ida2} we show the expected statistical error on $\tanb$ for 
$\mha = 100, 150$ and 200~GeV, assuming that the only measured
process is $\bb\ha$.  At the two higher $\mha$ values, in the MSSM context
it is the $\hh$ that would be decoupled and have mass
$\mhh\sim\mha$ and the $\hl$
would be SM-like.  Consequently, the $\bb\hh$ rate would be 
essentially identical to the $\bb\ha$ rate
and, assuming that one could verify the MSSM Higgs context by independent
means, would lead to still smaller $\tanb$ statistical errors than plotted,
the exact decrease depending upon the signal to background ratio.
For $\mha=150$ and $200\gev$, the $\hh\ha$ process (like the $\hl\ha$
process at $\mha=100\gev$) would have to
be computed in a specific model context or its relative weight
fitted by studying $\bb\bb$ production in greater detail in order
to minimize any systematic error from this source.

\begin{figure}[tp]
\caption{\label{fig:compare} 
For the MSSM with $\mha = 200$~GeV, and 
assuming $\call=2000\fbi$ at $\rts=500\gev$,
we plot the $1\sigma$ statistical error band in $\Delta\tanb/\tanb$
as a function of $\tanb$ based on:
(a) the rate for $\epem\to\bb\ha+\bb\hh\to\bb\bb$ (with the $\hh\ha$ pair
process reduced by the event selection); 
(b) the rate for $\epem\to\hh\ha\to\bb\bb$;
(c) the average of $\gamhhtot$ and $\gamhatot$ as determined
in $\epem\to\hh\ha\to\bb\bb$ events. Results for (a), (b) and (c)
all include running $b$-quark mass effects and employ
HDECAY \cite{hdecayref}.}
\begin{minipage}{0.7\textwidth}
\begin{center}
\hspace*{-.5in}
\includegraphics[width=.85\textwidth,angle=90]{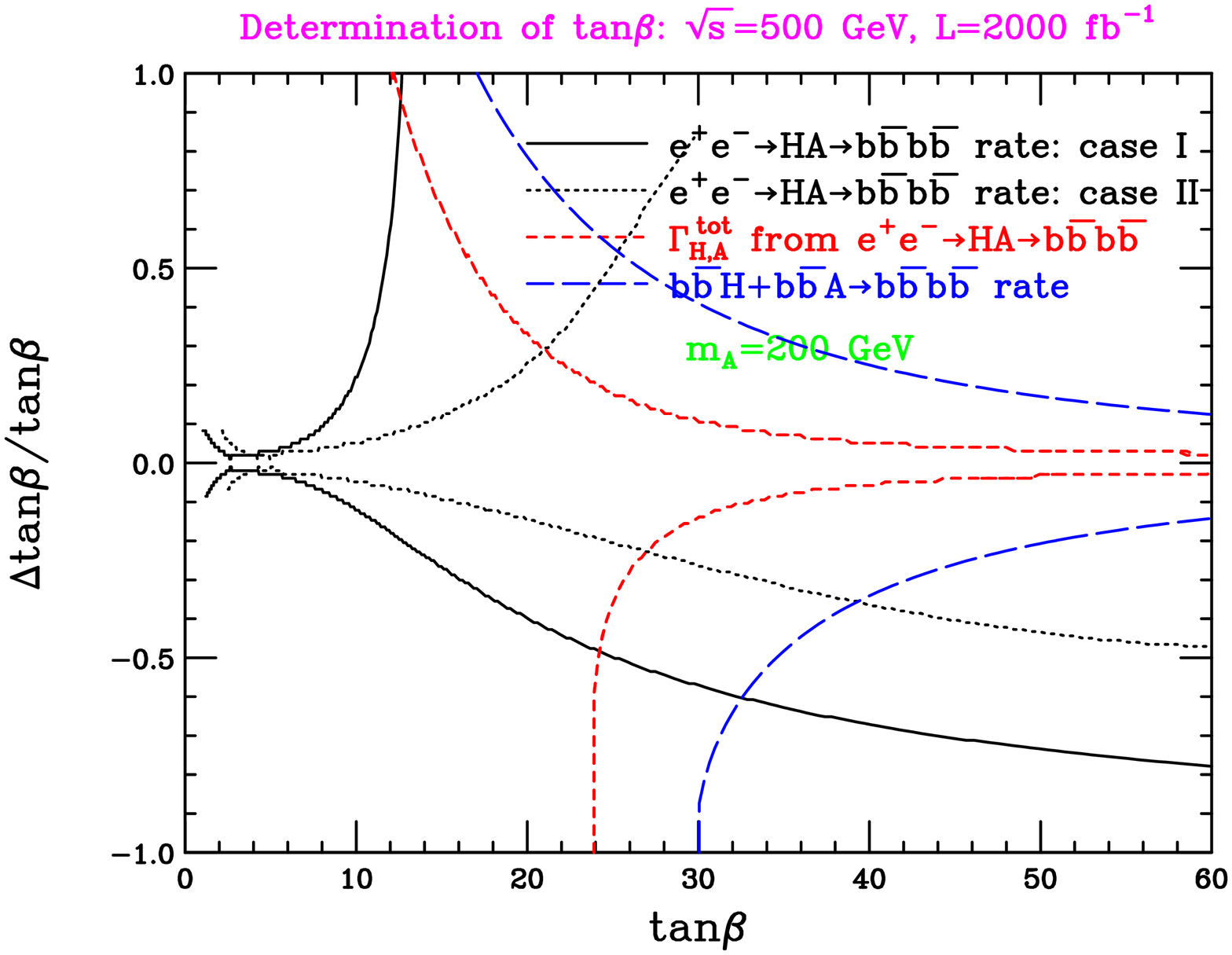}
\end{center}
\end{minipage}
\vspace*{-.7cm}
\end{figure}

\vspace*{-.6cm}
\section{\boldmath Complementary Methods:
$\hh,\ha$ branching ratios and total widths}
\vspace*{-4mm}

Owing to the large variation of the $\hh$, $\ha$ and $\hpm$ 
branching fractions to various allowed modes
 for low to moderate $\tanb$ in the MSSM,
$\tanb$ can be determined with good precision in this range
using $\hh\ha$ and $\hp\hm$ pair production (the cross sections
for which are nearly $\tanb$ independent). 
This was first demonstrated in~\cite{Gunion:1996cc,Gunion:1996qd}.
There, a number of models for which SUSY decays of the $\hh$,
$\ha$ and $\hpm$ are kinematically allowed were considered.
It was found that by measuring all available ratios of branching
ratios it was possible to determine $\tanb$ 
to better (often much better) than 10\% for $\tanb$ values 
ranging from 2  up to as high as 25 to 30
for $\mha$ in the 200--$400\gev$ range, assuming $\rts=1\tev$
and $\call_{\rm eff}=80\fbi$ (equivalent to $\call=2000\fbi$ for a
selection efficiency of 4\%).
A more recent analysis using a few specific points
in MSSM parameter space, focusing on the $\bb\bb$ event rate
and including a study at $\rts=500\gev$, is given in \cite{tao}.
This latter study uses a selection efficiency of 13\% 
and negligible background for detection of
$\epem \ra \hl\ha\to \bb\bb$ (relevant for $\mha\leq 100\gev$) 
or $\epem\to \hh\ha\to \bb\bb$ (relevant for $\mha\geq 150\gev$)
and finds small errors for $\tanb$ at lower $\tanb$ values.
Both \cite{Gunion:1996cc,Gunion:1996qd} and
\cite{tao} assume MSSM scenarios in which there are significant
decays of the $\ha$ and $\hh$ to pairs of SUSY particles, in particular
neutralinos and charginos.  These decays remain non-negligible
up to fairly high $\tanb$ values, as a result of which the $\bb$ branching
fractions of the $\ha$ and $\hh$ continue to vary noticeably as
$\tanb$ increases rather than being nearly constant. In the absence
of SUSY decays, the $\bb\bb$ rate would asymptote quickly to a fixed value
as $\tanb$ increases. As we shall see, this means that smaller errors
for the $\tanb$ determination using the $\hh\ha\to\bb\bb$ rate are achieved
if SUSY decays are present.

For this report we re-examined the errors on $\tanb$ that could be
achieved following procedures related to those of 
\cite{Gunion:1996cc,Gunion:1996qd,tao}, but 
using updated luminosity expectations and somewhat more realistic
experimental assumptions and analysis techniques. 
We restricted the analysis to the process $\epem\to\hh\ha\to \bb\bb$,
ignoring possible additional sensitivity through ratios relative
to other final states. With both Higgs bosons reconstructed in their $\bb$
final state as two back-to-back clusters of similar mass,
backgrounds are expected to be negligible.

Figure~\ref{fig:compare} compares the results for $\Delta\tanb/\tanb$
obtained using the $\epem\to\hh\ha\to \bb\bb$ rate to those based
on the $\bb\hh+\bb\ha\to \bb\bb$ rate (after including $b$-quark mass
running).  For the former, two different MSSM scenarios are considered:
\begin{description}
\itemsep=0.0in 
\item{(I)} $m_{\wt g}=1\tev$, $\mu=M_2=250\gev$, 
$m_{\wt t_L}=m_{\wt b_L}=m_{\wt t_R}=m_{\wt b_R}\equiv m_{\wt t}=1\tev$, 
$A_b=A_\tau=0$, $A_t=\mu/\tanb+\sqrt6 m_{\wt t}$ (maximal mixing);
\item{(II)}  $m_{\wt g}=350\gev$, $\mu=272\gev$, $M_2=120\gev$, 
$m_{\wt t_L}=m_{\wt b_L}=356\gev$, $m_{\wt t_R}=273\gev$, 
$m_{\wt b_R}=400\gev$, $A_\tau=0$, $A_b=-672\gev$, $A_t=-369\gev$.
\end{description}
In scenario (I), SUSY decays of the $\hh$ and $\ha$ are kinematically
forbidden. Scenario (II) is taken from \cite{tao} in which
SUSY decays (mainly to $\wt \chi_1^0\wt\chi_1^0$) are allowed.
In computing the statistical errors in $\tanb$, we assume 
an event selection efficiency of $10\%$ and no background;
$N(\bb\bb)\pm\sqrt{N(\bb\bb)}\geq 10$ is required
to set an upper (lower) $\tanb$ limit, respectively.
To give an idea of the sensitivity of the $\bb\bb$ event rate to $\tanb$,
we give a few numbers (assuming $\rts=500\gev$
and $\call=2000\fbi$);
the $\bb\bb$ event rate, after $10\%$ selection efficiency, 
is $1$, $5$, $34$, $1415$ $1842$ [$8$, $77$, $464$, $1762$, $1859$]
at $\tanb=1$, $2$, $3$, $10$, $40$, in scenarios (II) [(I)],
respectively.
These differing $\tanb$ dependencies imply significant 
sensitivity of the $\tanb$ errors to the scenario choice,
with worse errors for scenario (I). Where plotted,
errors for $\tanb$ from the $\bb\hh+\bb\ha\to\bb\bb$ rate are 
essentially independent of the scenario choice.

Regarding the $\tanb$ error from the $\hh\ha\to \bb\bb$ rate, 
we see from the above event numbers for scenario (I)
that once $\tanb$ reaches 10 to 12
the $\bb\bb$ rate will not change much if
$\tanb$ is increased further since the branching ratios are asymptoting.
In contrast, if $\tanb$ is decreased the $\bb\bb$ rate declines significantly
as other decay channels come into play. Thus, meaningful lower bounds
on $\tanb$ are retained out to relatively substantial $\tanb$ values
whereas upper bounds on $\tanb$ disappear for $\tanb\gsim 10-12$.
In scenario (II), 
we note that $\mhh$ begins to decrease for $\tan\beta\gsim 30$, resulting
in an increased $HA$ production cross section, which improves the
$\tan\beta$ limit.  However, there are significant theoretical uncertainties
in this region, and we cut off the curve at $\tan\beta=30$.
Obviously, the $\bb\hh+\bb\ha\to\bb\bb$ rate determination quickly becomes
far superior once $\tanb\gsim 20$.

Let us now turn to determining $\tanb$ using the intrinsic total
widths of the $\hh$ and $\ha$. Very roughly, it is only for $\tanb>10$
that they can provide a $\tanb$ determination. This is because (a)
the widths are only $> 5\gev$
(the detector resolution discussed below) for $\tanb> 10$ and (b) 
the number of events in the $\bb\bb$ final state becomes maximal once
$\tanb>10$. We first discuss the
experimental issues in determining the Higgs boson width. 
The expected precision of the SM Higgs boson width 
determination at the LHC and at a LC was studied~\cite{volker}. 
The statistical method used in~\cite{volker}
was based on a convolution of the estimated $\Gamma_{\rm res}=5$ GeV 
detector resolution
with a Breit-Wigner for the intrinsic width. It was applied to a $\hh\ha$
simulation~\cite{troncon} for a LC. An overall fit 
to the $b\anti b$ mass distribution gives a Higgs boson width 
which is about $2\sigma$ larger than expected from the convolution
of the 5 GeV resolution with the intrinsic Higgs width.
This can be traced to the fact that the overall fit includes
wings of the mass distribution that are present due to 
wrong pairings of the $b$-jets. The mass distribution contains
about 400 di-jet masses (2 entries per $\hh\ha$ event), of which about 
300 are in a central peak.  If one fits only the central peak,
the width is close to that expected based on
simply convoluting the 5 GeV resolution with the intrinsic Higgs width.
This indicates that about 25\% of the time wrong jet-pairings 
are made and contribute to the wings of the mass distribution.
Therefore, our estimates of the error on the
determination of the Higgs width will be based on the assumption that
only 3/4 of the events ({\it i.e.} those in the central peak)
retained after our basic event selection cuts (with assumed
selection efficiency of $10\%$)
can be used in the statistics computation.
The $m_{\bb}$ for each of the $\bb$
pairs identified with the $\hh$ or $\ha$ is binned in a single 
mass distribution. This is appropriate
since the $\hh$ and $\ha$ are highly degenerate for
the large $\tanb$ values being considered. Thus, our observable
is the average of the widths $\gamhhtot$ and $\gamhatot$.
Finally, we
note that the detector resolution will not be precisely determined.
There will be a certain level of systematic uncertainty
which we have estimated at 10\% of $\Gamma_{\rm res}$, {\it i.e.} 0.5 GeV.
This systematic uncertainty considerably weakens our ability to
determine $\tanb$ at the lower values of $\tanb$ for which
$\gamhhtot$ and $\gamhatot$ are smaller than $\Gamma_{\rm res}$. This
systematic uncertainty
should be carefully studied as part of any eventual experimental analysis.

Our study is done in the context of the MSSM 
and assumes the stated soft SUSY breaking
parameters. For these parameters, the one-loop corrections to 
the $b\anti b$ couplings of the $\hh$ and $\ha$ and the 
stop/sbottom mixing present in the one-loop corrections to the Higgs
mass matrix \cite{Carena:1998gk} are small.  More generally,
however, substantial ambiguity can arise
if the sign and magnitude of $\mu$ is not fixed.
However, assuming that these parameters are known,
the results for the error on $\tanb$ from the width
measurement are quite insensitive to the precise scenario.  Indeed,
results for our two SUSY scenarios (I) and (II) are indistinguishable.

The resulting accuracy for $\tanb$ obtained from measuring the average
$\hh/\ha$ width is shown in Fig.~\ref{fig:compare}, assuming
$\mha=200\gev$, $\call=2000\fbi$ and $\rts=500\gev$. We see that good
accuracy is already achieved for $\tanb$ as low as 25 with
extraordinary accuracy predicted for very large $\tanb$.  The sharp
deterioration in the lower bound on $\tanb$ for $\tanb\lsim 24$ occurs
because the width falls below $\Gamma_{\rm res}$ as $\tanb$ is taken
below the input value and sensitivity to $\tanb$ is lost.  If there
were no systematic error in $\Gamma_{\rm res}$, this sharp fall off
would occur instead at $\tanb\lsim 14$.  To understand these effects
in a bit more detail, we again give some numbers for scenario (II). At
$\tanb=50$, $55$ and $60$, 
$\vev{\gamhhtot,\gamhatot}\sim 10.4$, $12.5$ and $14.9\gev$, 
respectively. After including the detector resolution, 
the effective average widths
become 11.5, 13.4 and 15.7 GeV, respectively, whereas the total
error in the measurement of the average width, including
systematic error, is $\sim 0.54 \gev$.  Therefore,
$\tanb$ can be determined to about $\pm 1$, or
to better than $\pm 2\%$. This high-$\tanb$ situation can
be contrasted with $\tanb=15$ and 20,
for which $\vev{\gamhhtot,\gamhatot}=0.935$ and 1.64 GeV, respectively,
which become 5.09 and 5.26 GeV after including detector resolution.
Meanwhile, the total error, 
including the statistical error and the 
systematic uncertainty for $\Gamma_{\rm res}$,
is about 0.57 GeV.

The accuracies from the
width measurement are somewhat better than those
achieved using the $\bb\ha+\bb\hh\to\bb\bb$ rate measurement. Of
course, these two high-$\tanb$ methods for determining $\tanb$ are
beautifully complementary in that they rely on very different
experimental observables. Both methods are nicely complementary in
their $\tanb$ coverage to the $\tanb$ determination based on the
$\hh\ha\to\bb\bb$ rate, which comes in at lower $\tanb$. Still, there
is a window,  $10\lsim \tanb\lsim25$ in scenario (I) or
$20\lsim \tanb\lsim 25$ in scenario (II), for which an accurate 
determination of $\tanb$ ($\Delta\tanb/\tanb<0.2$) 
using just the $\bb\bb$ final state processes will not be possible.  
This window expands rapidly as $\mha$ increases (keeping $\rts$ fixed).
Indeed, as $\mha$ increases above $250\gev$, $\hh\ha$ pair production
becomes kinematically forbidden at $\rts=500\gev$ and detection of the
$\bb\hh+\bb\ha$ processes at the LC (or the LHC) requires
\cite{Grzadkowski:1999wj} increasingly large values of $\tanb$.  This
difficulty persists even for $\rts\sim 1\tev$ and above; if
$\mha>\rts /2$, the $\hh$ and $\ha$ cannot be pair-produced and yet
the rate for $\bb\hh+\bb\ha$ production is undetectably small 
for moderate $\tanb$ values.

In the above study, we have not made use of other decay channels of
the $\hh$ and $\ha$, such as $\hh\to WW,ZZ$, $\hh\to \hl\hl$, $\ha\to
Z\hl$ and $\hh,\ha\to$SUSY.  As in the studies of
\cite{Gunion:1996cc,Gunion:1996qd}, their inclusion should
significantly aid in determining $\tanb$ at low to moderate $\tanb$
values.  A determination of $\vev{\gamhhtot,\gamhatot}$ is also
possible using the $\bb\ha+\bb\hh\to\bb\bb$ events.  Assuming that
50\% of the events selected in the analysis of Section II can be used
for a fit of the average width and that $5\gev$ resolution with 10\%
systematic error for the width measurement can be achieved, the
resulting $\tanb$ errors are similar to those from the
$\bb\ha+\bb\hh\to\bb\bb$ event rate for $\tanb>30$. A complete
analysis that takes into account the significant background and the
broad energy spectrum of the radiated $\hh$ and $\ha$ is needed.
However, it should be noted that this is the only width-based
technique that would be available if $\hh\ha$ pair production is not
kinematically allowed.  We have also not employed charged Higgs boson
production processes. In $\epem\to\hp\hm$ production, the absolute
event rates and ratios of branching ratios in various channels will
increase the $\tanb$ accuracy at low $\tanb$
\cite{Gunion:1996cc,Gunion:1996qd,feng} and the total $\hpm$ width
measured in the $tb$ decay channel will add further precision to the
$\tan\beta$ measurement at high $\tanb$.  The rate for $\epem\to
t\anti b H^-+\anti t b H^+ \to t\anti t b\anti b$ is also very
sensitive to $\tanb$ and might be a valuable addition to the
$\epem\to\bb\ha+\bb\hh\to \bb\bb$ rate determination of $\tanb$.  The 
theoretical study of
\cite{feng} finds, for example, that if $\mhpm=200\gev$ and
$\tanb=50$ ($\tanb=20$), then the $1\sigma$ errors (including systematic
uncertainties) on $\tanb$ 
are $\Delta\tanb/\tanb=0.06$ ($\Delta\tanb/\tanb=0.2$), respectively,
for $\call=2000\fbi$ and $\rts=500\gev$.

\vspace*{-4mm}
\section{Conclusions}
\vspace*{-2mm}

A high-luminosity linear collider is unique in its
ability to precisely measure the value of $\tanb$. 
This is because highly precise measurements of Higgs boson production
processes will be essential and are only possible at the LC. 
%
In the context of the MSSM, a variety
of complementary methods will allow an accurate determination of $\tanb$ 
over much of its allowed range, including, indeed especially for, 
large $\tanb$ values, provided $\mha\lsim \rts/2$. 
In particular, we have demonstrated the complementarity
of employing: a) the rate for $\bb\ha+\bb\hh \to \bb\bb$; b) 
the $\hh\ha\to \bb\bb$ rate; and c) 
a measurement of the average $\hh,\ha$ total width in $\hh\ha$ production. 
The analogous charged
Higgs observables --- the $tb\hpm\to tb tb$ rate,
the $\hp\hm\to t\anti b \anti tb$ rate 
and the total $\hpm$ width measured in $\hp\hm$ production --- will
further increase the sensitivity to $\tanb$.
The possible impact of MSSM radiative corrections on
interpreting these measurements \cite{Carena:1998gk}
will be discussed in a longer note.
In the general 2HDM, if, for example, the only non-SM-like Higgs
boson with mass below $\rts$ is the $\ha$, then a good determination
of $\tanb$ will be possible at high $\tanb$ from the 
$\bb\ha\to\bb\bb$ production rate.

\bigskip
\centerline{\bf Acknowledgments}
\smallskip
This work was supported in part by the U.S. Department of Energy,
the Davis Institute for High Energy Physics and the Wisconsin Alumni
Research Foundation.


\end{document}